\documentstyle[aps,multicol,tabularx,rotate,epsf]{revtex}
\begin{document}

\title{Casimir Torques between Anisotropic Boundaries in Nematic Liquid
Crystals} 

\author{Ramin Golestanian$^{1,2}$, Armand Ajdari$^{1}$, and
Jean-Baptiste Fournier$^{1}$} 

\address{$^{1}$Laboratoire de Physico-Chimie Th\'eorique, ESA CNRS 7083,
E.S.P.C.I., 75231 Paris Cedex 05, France\\ $^{2}$Institute for Advanced
Studies in Basic Sciences, Zanjan, 45195-159, Iran}

\date{\today}

\maketitle
\begin{abstract}
Fluctuation-induced interactions between anisotropic objects immersed in
a nematic liquid crystal are shown to depend on the relative orientation
of these objects. The resulting long-range ``Casimir'' torques are
explicitely calculated for a simple geometry where elastic effects are
absent. Our study generalizes previous discussions restricted to the
case of isotropic walls, and leads to new proposals for experimental
tests of Casimir forces and torques in nematics.

\medskip
\noindent Pacs numbers: 68.60.Bs, 61.30.Hn, 61.30.Dk
\end{abstract}

\begin{multicols}{2}

In the last decade, much theoretical attention has been paid to
``Casimir'' forces in structured complex
fluids~\cite{KGRMP,Krech,Mostepa}. The pioneering work of Casimir showed
that two uncharged conducting plates attract each other in vacuum, due
to the modification of the electromagnetic fluctuations imposed by the
plates~\cite{Casimir}. In complex fluids, similar interactions should
exist between embedded objects, as the thermal fluctuations
of the medium's elastic distortions are restrained by the boundary
conditions imposed by the objects.  These interactions are believed to
act between bounding surfaces or immersed inclusions in critical fluids
or superfluids~\cite{Symanzik,LiK}, in liquid
crystals~\cite{Armand,Denis,Ziherl1,Ziherl2}, in bilayer
membranes~\cite{Goulian,GGK,Dommersnes}, and also between rodlike
polyelectrolytes~\cite{oosawa,BJ,Pod,HaLiu}.  Nematic liquid crystals
are anisotropic fluids with a quadrupolar long-range order. They are
considered as good candidates for the direct observation of ``Casimir''
interactions in complex fluids.  However, clear experimental evidences
have to date been scarce~\cite{Krech}. This is probably due to the
weakness of fluctuation-induced effects when compared to that of
permanent elastic distortions, which are often present.

Although nematic liquid crystals are well-known to display
orientational effects~\cite{deGennes}, no ``Casimir'' interaction
directly connected to these orientational properties of nematics
have so far been discussed.  Here, we demonstrate that thermal
fluctuations in nematic liquid crystals can induce {\em torques}
between bounding surfaces~\cite{electro}. The existence of ``Casimir'' torques
between objects embedded in complex fluids is usually caused by
the anisotropy in the {\em shape\/} of the
objects~\cite{GGK,BJ,HaLiu}; here we report on a more subtle
effect occurring between infinite plates with translational
symmetry. To emphasize the ``Casimir'' effect, we focus on a
situation in which no average elastic distortion is present: two
parallel plates with a surface energy favoring an orientation of
the local average molecular alignment (director) {\em
perpendicular\/} to the surfaces. The ground state is therefore
the distorsionless state in which the director is everywhere
perpendicular to the boundaries. A ``Casimir'' torque can arise
due to the {\em anisotropy} in the rigidity of the surface energy:
we assume that deviations from the preferred normal surface
orientation is easier in one direction than in the orthogonal one.
(This property can be experimentally obtained from a grating surface
treated for homeotropic anchoring~\cite{BryanBrown98}, or by depositing
on top of a substrate which is conventionally treated to give planar
anchoring a very thin layer of a material promoting homeotropic
anchoring~\cite{Barberi}.) In this geometry, we calculate the
``Casimir'' interaction, and show that it depends not only on the
distance between the two plates, but also on their relative orientation.
At the end of the paper, we argue that this leads to effects easier to
measure than the direct force between the plates.
 
Nematic liquid crystals are liquids of rodlike molecules
displaying a long-range orientational order. The local average
molecular axis is represented by a unit vector $\pm\bf n$, called
the ``director''.  The bulk ground state corresponds to a uniform
director field and the Frank elasticity describes the free energy
associated with gradients of the director~\cite{Frank}. Bounding
plates often favor some orientation of the director: this
phenomenon is known as {\em anchoring\/}~\cite{deGennes}. The
simplest situations correspond to a preferred orientation normal
to the plates (homeotropic anchoring) or parallel to the plates
(planar anchoring). Here we employ a path integral method to study
the ``Casimir'' energy of a nematic liquid crystal confined
between two parallel plates at a separation $d$, at which we
assume an homeotropic anchoring with anisotropic strength as
described above. We calculate the fluctuation-induced interaction
between the two plates when the axes of weakest anchoring strength
are placed at a relative angle $\theta$ (see Fig.~1), and find
\begin{equation}
{\cal F}(\theta,d)=
-\frac{k_{\rm B}T}{8\pi d^2}\times
\sum_{k=1}^{\infty}
{\cos^k 2 \theta \over k^3},
\label{Ftheta}
\end{equation}
per unit area of the plates. The angle dependence in this interaction
shows that the plates experience a long-range fluctuation-induced torque,
that decays algebraically as $1/d^2$, and
tends to align the
directions of weakest rigidity.

We start with the broken symmetry configuration of lowest energy,
in which the director field is aligned along the $z$-axis, and
restrict ourselves to small fluctuations $\delta {\bf n}={\bf
n}-\hat{\bf z}$ of the director around this ground state (see
Fig.~1). The bulk cost of such a fluctuation can be described by
an effective Hamiltonian
\begin{equation}
{\cal H}=\frac{K}{2}\int\!\!\!\int\!\!\!\int\!d^3{\bf r}\left[\nabla \delta
{\bf n}({\bf r}) \right]^2,  \label{H}
\end{equation}
which corresponds to the usual one-constant ($K$) approximation of the
Frank elasticity~\cite{Frank,deGennes}. The anisotropic homeotropic anchoring
surface energy is accounted for by the surface Hamiltonian
\begin{equation}
{\cal H}_s={1 \over 2} \int\!\!\!\int\!d^2{\bf r}_\perp\left( \delta
{\bf n}^-\!\cdot
{\sf W}^-\!\cdot\delta {\bf n}^- + \delta {\bf n}^+\!\cdot{\sf
W}^+\!\cdot\delta
{\bf n}^+ \right), \label{Hs}
\end{equation}
where $\delta{\bf n}^-({\bf r}_\perp)=\delta{\bf n}({\bf r}_\perp,0)$
and $\delta {\bf n}^+({\bf r}_\perp)=\delta {\bf n}({\bf r}_\perp,d)$,
and ${\sf W}^-$ and ${\sf W}^+$ are positive definite constant tensors that entail
the anisotropy of the surfaces and their relative orientation. We assume
that the two surfaces are identical in nature and we define
${\sf W}^-=W_{\rm max} \,{\hat{\bf x}}^-\!\otimes{\hat{\bf x}}^- + W_{\rm
min}\,{\hat {\bf y}}^-\!\otimes{\hat{\bf y}}^-$ and ${\sf W}^+=W_{\rm
max}\,{\hat {\bf x}}^+ \!\otimes{\hat {\bf x}}^+ + W_{\rm min}\,{\hat
{\bf y}}^+\!\otimes{\hat {\bf y}}^+$, where ${\hat {\bf x}}^\pm$ and
${\hat {\bf y}}^\pm$ denote the hard and weak axis on plate $\pm$,
respectively (see Fig.~1). The eigenvalues $W_{\rm max}$ and $W_{\rm
min}$ naturally define two extrapolation lengths $\lambda_{\rm
min}={K/W_{\rm max}}$ and $\lambda_{\rm max}={K/W_{\rm min}}$
\cite{deGennes}. We assume extreme anisotropy, namely $\lambda_{\rm min} \ll
\lambda_{\rm max}$.

\begin{figure}
\centerline{\epsfxsize=5cm\epsfbox{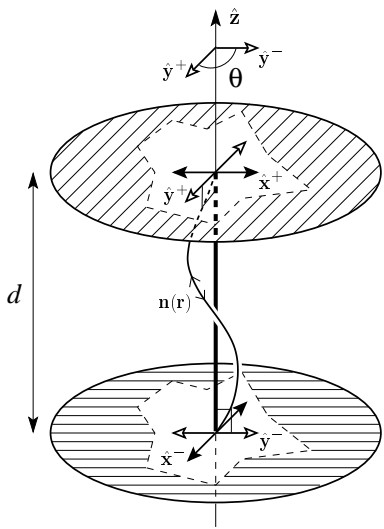}}
\caption{Sketch of a fluctuation of the bulk director field ${\bf n}({\bf
r})$  (thin line) around its mean value $\hat{\bf z}$ (bold line).
The two anisotropic
plates impose a homeotropic anchoring (i.e., the preferred
surface value of ${\bf n}$ is $\hat{\bf z}$) characterized by
an anisotropic anchoring elasticity: deviations from $\hat{\bf z}$ are easier
in the ``soft'' directions $\hat{\bf y}^\pm$ than in the ``hard''
directions $\hat{\bf x}^\pm$.
In this geometry, the fluctuation-induced ``Casimir'' interaction between the
plates depends both on their separation~$d$ and
their relative angle~$\theta$.}
\end{figure}

``Casimir'' effects arise from the quantization of the fluctuation
modes by the boundaries (essentially from the suppression of modes
of wavevector smaller than $2\pi/d$). The actual effect of the
boundaries on a fluctuation mode of wavevector $q\simeq 2\pi/d$ is
a function of the product $q\lambda$, where $\lambda$ is the
extrapolation length corresponding to the considered polarization
of the fluctuation. Depending on the relative values of
$\lambda_{\rm min}$, $\lambda_{\rm max}$ and $d$, three different
regimes occur: (i) For $\lambda_{\rm min}\ll\lambda_{\rm max}\ll
d$ the constraints are effectively hard in both directions and the
anisotropy is washed out at leading order. (ii) For $\lambda_{\rm
min}\ll d\ll\lambda_{\rm max}$ the director field is subject to a
hard constraint in the $x$ direction while it is virtually free in
the (perpendicular) $y$ direction. (iii) Finally, for
$d\ll\lambda_{\rm min}\ll\lambda_{\rm max}$ both directions allows
almost free fluctuations and the anisotropy is lost again at
leading order. To emphasize the effect of the anisotropy, we thus
focus on the case where $\lambda_{\rm min} \ll d\ll\lambda_{\rm
max}$.

Following Ref.~\cite{Armand}, we now proceed to calculate the partition
function for the director fluctuations, using ${\cal H}+{\cal H}_s$ as
the total Hamiltonian.  The quantum mechanical description of the
partition function \cite{Armand}, which treats $z$ as an imaginary time
variable, helps us gain a useful insight into the meaning of the
boundary conditions. From the ``imaginary time action'' ${\cal H}$, one
can define the ``momentum'' conjugate to a component of $\delta{\bf n}$
as $P_{\delta n}=K\,\partial_z \delta n$. Then, one can go from
coordinate space to momentum space. In particular, at the boundaries one
finds out that $P_{\delta n}$ is quadratically confined by the tensor
${\sf W}^{-1}$ and thus acts opposite as compared to $\delta n$. In
other words, there is an uncertainty principle relating the fluctuations of
$\delta n$ and $\partial_z \delta n$ as $\Delta(\delta n)
\Delta(\partial_z \delta n) \simeq k_{\rm B} T/K$. In light of this, one
can argue that the boundary condition of nearly free fluctuations in the
soft direction is asymptotically equivalent to setting $\partial_z
\delta n=0$. Note that this is equivalent to assuming that the directors
cannot bear torques due to the freedom of rotation in the soft
direction.

With the above justification, we can employ a somewhat less involved
path integral formulation \cite{KGRMP,LiK} using the following
boundary conditions: $\delta{\bf n}^\pm\cdot{\hat {\bf x}}^\pm=0$,
$\partial_z \delta {\bf n}^\pm\cdot {\hat {\bf y}}^\pm =0$. The
partition function of the fluctuating director field subject to the
above constraints can be written as
\begin{eqnarray}    \label{Z1}
{\cal Z}&=&\int\!{\cal D} \delta {\bf n}({\bf r})\,\,\delta\{\delta
{\bf n}^- \cdot {\hat {\bf x}}^-\} \,\delta\{\partial_z \delta
{\bf n}^- \cdot {\hat {\bf y}}^-\} \\
&& \times \, \delta\{\delta {\bf n}^+ \cdot {\hat {\bf x}}^+\} \,
\delta\{\partial_z \delta {\bf n}^+ \cdot {\hat {\bf x}}^+\} \,
e^{-{\cal H}/{k_{\rm B} T}}. \nonumber
\end{eqnarray}
The functional delta functions can be written as integral
representations by introducing four Lagrange multiplier surface fields:
\end{multicols}
\begin{eqnarray}    \label{Z2}
{\cal Z}&=& \int\!{\cal D}\delta{\bf n}({\bf r})\,{\cal D}
l_1({\bf r}_\perp)\,{\cal D}l_2({\bf r}_\perp)\,{\cal D}l_3({\bf
r}_\perp)\,{\cal D}
l_4({\bf r}_\perp) \,\,
\exp\left\{-{\kappa\over 2} \int\!\!\!\int\!\!\!\int\!d^3{\bf r}
\left[\nabla \delta {\bf n}({\bf r}) \right]^2 \right. \\
&&+\left.i \int\!\!\!\int\!d^2{\bf r}_\perp \left[
l_1({\bf r}_\perp)\,\delta{\bf n}^-({\bf r}_\perp)\cdot{\hat {\bf x}}^-
+
l_2({\bf r}_\perp)\,\partial_z\delta{\bf n}^-({\bf r}_\perp)
\cdot{\hat {\bf y}}^-
+
l_3({\bf r}_\perp)\,\delta{\bf n}^+({\bf r}_\perp)\cdot{\hat {\bf x}}^+
+
l_4({\bf r}_\perp)\,\partial_z\delta{\bf n}^+({\bf r}_\perp)
\cdot{\hat {\bf y}}^+
\right]\right\},
\nonumber
\end{eqnarray}  \noindent
where $\kappa=K/k_{\rm B} T$. The integration over the director
field is now Gaussian and can be easily performed. It yields
\begin{equation}    \label{Z3}
{\cal Z}=\int\!{\cal D} l_1({\bf q}_\perp)
\,{\cal D} l_2({\bf q}_\perp)\,{\cal D} l_3({\bf q}_\perp)\,{\cal D} l_4({\bf q}_\perp)\,
\exp\left\{-\frac{1}{4\kappa}\!\sum_{\alpha,\beta=1}^{4}\!\int\!\!\!\int\!  {d^2
{\bf q}_\perp \over
(2\pi)^2} \, l_\alpha(-{\bf q}_\perp)\,M_{\alpha\beta}({\bf q}_\perp)\,
l_\beta({\bf q}_\perp) \right\},
\end{equation}
in which
\begin{equation}
{\sf M}(q)=\left[\begin{array}{cccc} \displaystyle{1 \over q} & 0 &
\displaystyle{e^{-q d} \over q} \cos\theta & -e^{-q d} \sin\theta \\ \\
0 & -q & e^{-q d} \sin\theta & -q e^{-q d}\cos\theta \\ \\
\displaystyle{e^{-q d} \over q} \cos\theta & e^{-q d} \sin\theta &
\displaystyle{1 \over q} & 0 \\ \\
-e^{-q d} \sin\theta &  -q e^{-q d} \cos\theta & 0 & -q
\end{array}\right], \label{Mq}
\end{equation}
\begin{multicols}{2}    \noindent
where $\theta$ is the angle between the corresponding soft axes of the
two plates (see Fig.~1). The remaining integration over the Lagrange
multiplier fields can be performed to give $\ln{\cal Z}=-{1 \over
2} \sum_{{\bf q}_\perp} \ln\left(\det {\sf M}({\bf q}_\perp)\right)$, which leads to
a simple expression for the free energy per unit area:
\begin{equation}
{\cal F}(\theta,d)={k_{\rm B} T \over 2 \pi d^2}
\int_{0}^{\infty}\! d u \, u \, \ln\left[1-e^{-2 u}
\cos2\theta\right]. \label{F2}
\end{equation}
Integration over $u$ then gives the final result Eq.~(\ref{Ftheta}) above. The function
$f(\theta)=\sum_{k=1}^{\infty} {\cos^k (2\theta) / k^3}$ which
describes the orientational dependence of the interaction has the
structure of a zeta function that is commonly present in ``Casimir''
interactions.

It is instructive to examine the limiting cases of plates in which the
corresponding hard and soft axes are parallel or perpendicular to each
other. For $\theta=0$ the boundary conditions corresponds to a hard-hard
configuration for one component of $\delta {\bf n}$ and a soft-soft one
for the other. One can check that $f(0)=\zeta(3)\simeq1.20206$ ($\zeta$
is the zeta function), thus we obtain exactly the same expression for
the ``Casimir'' energy as in Ref.~\cite{Armand} for ``alike'' boundary
conditions. On the other hand $f({\pi \over 2})=-{3 \over 4} \zeta(3)$,
which gives the same result as in Ref.~\cite{Armand} for ``unlike''
boundary conditions ($\theta=\pi/2$ corresponds to a soft-hard
configuration for both components of $\delta {\bf n}$).

Our calculation suggests two kind of experiments: (i) a direct measure of the
torque exerted between two plates at a fixed separation, and (ii) a
measure of force as a function of separation for plates at various
angles.

A possible experimental setup for the direct observation of the
fluctuation-induced torque could be a torsion pendulum similar to the
one discussed in Ref.~\cite{exp}. Our results imply that the torsion
coefficient of the pendulum $k$ (defined as the ratio between the torque
$\tau$ and the angular rotation $\Delta\theta$) is corrected at
$\theta=0$ by an amount
\begin{equation}
\Delta k={\pi^2 \over 12 } \times {k_{\rm B} T R^2 \over
d^2},\label{dk}
\end{equation}
due to the ``Casimir'' effect. Here $R$ is the radius of the plates of
area $\pi R^2$. Using the typical values $k_{\rm B} T=4.1\times
10^{-14}$~dyn\,cm, $R=1.5$~cm, $d=10^{-4}$~cm, one obtains $\Delta k \sim
10^{-5}$~dyn\,cm. This accuracy may be reachable using modern
micro-manipulation techniques.

A measure of the force-distance relation for various angles
$\theta$ could also be performed. An advantage of this procedure,
as compared to measurement of the ``simpler'' effect corresponding
to isotropic anchoring, is that relative effects are more easily
detectable.  Indeed, while the weak signal of the ``Casimir''
force can be swamped by stronger effects (Van der Waals, etc.),
the difference between measurements performed at $\theta =0$ and
$\theta=90^\circ$ should provide a differential evidence of the
Casimir scaling.


We are grateful to R.~Barberi, I.~Dozov, P.~Galatola and L.~Peliti for
invaluable discussions and comments. This research was supported in part
by the National Science Foundation under Grant No.\ DMR-98-05833, and by
ESPCI through a Joliot visiting chair for one of us (RG).

\end{multicols}

\begin{references}

\bibitem{KGRMP}
For a review, see M. Kardar and R. Golestanian, Rev.\ Mod.\ Phys.\ {\bf
71}, 1233 (1999).

\bibitem{Krech}
M. Krech, {\em The Casimir Effect in Critical Systems} (World
Scientific, Singapore, 1994).

\bibitem{Mostepa}
V. M. Mostepanenko and N.N. Trunov, {\em The Casimir Effect and Its
Applications} (Clarendon Press, Oxford, 1997).

\bibitem{Casimir}
H. B. G. Casimir, Proc.\ Kon.\ Ned.\ Akad.\ Wetenschap B {\bf 51}, 793
(1948).

\bibitem{Symanzik}K. Symanzik, Nucl.\ Phys.\ {\bf B190}, 1 (1981).

\bibitem{LiK}
H. Li and M. Kardar, Phys.\ Rev.\ Lett.\ {\bf 67}, 3275 (1991);
Phys.\ Rev.\ A {\bf 46}, 6490 (1992).

\bibitem{Armand}
A. Ajdari, L. Peliti, and J. Prost, Phys.\ Rev.\ Lett.\ {\bf 66},
1481 (1991); A. Ajdari, B. Duplantier, D. Hone, L. Peliti, and J.
Prost, J. Phys.\ II {\bf 2}, 487 (1992).

\bibitem{Denis}
D. Bartolo, D. Long and J.-B. Fournier.  Europhys.\ Lett.\ {\bf 49}, 729
(2000).

\bibitem{Ziherl1}P. Ziherl, R. Podgornik, and S. Z{\v u}mer, Phys.\
Rev.\ Lett.\ {\bf 82}, 1189 (1999).

\bibitem{Ziherl2}P. Ziherl, F. K. P. Haddadan, R. Podgornik, and S. Z{\v
u}mer, Phys.\ Rev.\ E {\bf 61}, 5361 (2000).

\bibitem{Goulian}
M. Goulian, R. Bruinsma, and P. Pincus, Europhys.\ Lett.\ {\bf 22}, 145
(1993).

\bibitem{GGK}
R. Golestanian, M. Goulian, and M. Kardar, Europhys.\ Lett.\ {\bf
33}, 241 (1996); Phys.\ Rev.\ E {\bf 54}, 6725 (1996); R.
Golestanian, Phys.\ Rev.\ E, {\bf 62}, 5242 (2000).

\bibitem{Dommersnes}
P. G. Dommersnes and J.-B. Fournier.  Europhys.\ Lett.\ {\bf 46}, 256
(1999); Eur.\ Phys.\ J. B {\bf 12}, 9 (1999).

\bibitem{oosawa}
F. Oosawa, {\it Polyelectrolytes} (Marcel Dekker, New York, 1971).

\bibitem{BJ}
J.-L. Barrat and J.-F. Joanny, Adv.\ Chem.\ Phys.\ {\bf XCIV}, 1
(1996).

\bibitem{Pod}
R. Podgornik and V. A. Parsegian, Phys.\ Rev.\ Lett.\ {\bf 80}, 1560
(1998).

\bibitem{HaLiu}
B.-Y. Ha and A. J. Liu, Europhys.\ Lett.\ {\bf 846}, 624 (1999).

\bibitem{deGennes}
P.-G. de Gennes and J. Prost, {\it The Physics of Liquid
Crystals} (Clarendon, Oxford, 1993).

\bibitem{electro}Note that similar ideas have been recently introduced
in the case of electromagnetic quantum fluctuations by O.~Kenneth and
S.~Nussinov, preprint hep-th/9802149; hep-th/0001045.

\bibitem{BryanBrown98}G. P. Bryan-Brown, C. V. Brown, I. C. Sage, and V.
C. Hui, Nature {\bf 392}, 365 (1998).

\bibitem{Barberi}R. Barberi, private communication.

\bibitem{Frank}F. C. Frank, Discuss.\ Faraday Soc.\ {\bf 25}, 19 (1958).

\bibitem{exp}
S. Faetti, M. Gatti, and V. Palleschi, J. Physique Lett. {\bf 46},
L-881 (1985).

\end{references}
\end{document}